\documentclass[12pt]{iopart} 
 \usepackage[]{graphicx} 

\newcommand{\be}{\begin{equation}} \newcommand{\ee}{\end{equation}}
\newcommand{\ba}{\begin{align}} \newcommand{\ea}{\end{align}}

\begin{document}

\title[Polynomial approximations for Bessel functions and fractional diffusion problems]{On a novel iterative method to compute  polynomial approximations to Bessel functions of the first kind and its connection to the solution of fractional diffusion/diffusion-wave problems}

\author{Santos Bravo Yuste and Enrique Abad}
\address{Departamento de F\'{\i}sica, Universidad  de  Extremadura, E-06071 Badajoz, Spain}

\begin{abstract} We present an iterative method to obtain approximations to Bessel functions of the first kind $J_p(x)$ ($p>-1$) via the repeated application of an integral operator to an initial seed function
$f_0(x)$.  The class of seed functions $f_0(x)$ leading to sets of increasingly accurate approximations
$f_n(x)$   is considerably large and includes any polynomial. When the operator is applied once to a polynomial of degree $s$, it yields a polynomial of degree $s+2$, and so the iteration of
this operator generates sets of increasingly better polynomial approximations of increasing degree. We focus on the set of polynomial approximations  generated from the seed function
$f_0(x)=1$. This set of polynomials is not only useful for the  computation of
$J_p(x)$, but also from a physical point of view, as it describes the long-time decay modes of certain fractional diffusion and diffusion-wave problems.

\end{abstract}

\date{\today}

\maketitle

\section{Introduction}

Bessel functions play a central role in numerous problems in Science \cite{Various, Crank, Carslaw-Jaeger, Arfken, Dutka}. In particular, Bessel functions of
the first kind $J_p(x)$  appear in the solution of many problems involving wave and heat equations,
especially in systems with spherical or cylindrical symmetry \cite{Crank,Carslaw-Jaeger,Arfken}. Such
functions have been intensively studied for more than three hundred years \cite{Dutka} and a large body
of results concerning their properties is now available \cite{Watson,AbramowitzStegun,WolframBessel}.
It is well-known that they can be expressed in differential and integral forms, in terms of infinite series,
as a solution of differential equations or recursive relations, etc. Given their ubiquity, their efficient
numerical computation and its approximation in terms of simpler functions is an issue of great interest
\cite{NumericalRecipes,Gross,MillaneEads,LiLiGross}.
In particular, the approximation of special functions in terms of polynomials has some key advantages, as the latter can be effortlessly evaluated and manipulated.

In what follows we show how to derive infinite \emph{sets} of
polynomial approximations to $J_p$ valid for $p>-1$  [and, of course, also for negative integer values of $p$ by virtue of the relation $J_{-p}(x)= (-1)^p J_p(x)$ for $p$ integer]. Each set is generated from an initial polynomial seed function  $f_0(x)$ via the iteration of an integral operator and it eventually converges towards
a properly normalized Bessel function of the first kind, $\tilde J_p(x)\equiv 2^p p!  J_p(z_p x)/(z_p x)^p$, where $z_p$ denotes the first zero of $J_p(x)$. When applied to a polynomial approximation of degree $s$, the integral operator yields an improved polynomial estimate of degree $s+2$. The set of polynomial approximations  ${\rm Ba}_n^{(p)} $ we primarily focus on stems from the possibly simplest seed function  $f_0(x)=1$. However, we also present some results for another set of polynomials ${\rm Be}_n^{(p)} $ corresponding to the choice $f_0(x)=1-x$.

Our method  is inspired by the interesting properties of the integral operator
\be \Lambda_{p}\left[ f \right]=z_p^2 \int_x^1\frac{du}{u^{2p+1}} \int_0^udv v^{2p+1} f(v) \ee
 recently introduced  by the authors in collaboration with Borrego \cite{divergentSeriesYBA} to study Fourier-Bessel solutions of fractional diffusion equations. The operator $\Lambda_{p}$ is closely related to Bessel's differential equation and it has the remarkable property of leaving the function  $\tilde J_p(x)$ invariant (cf.~\ref{sec:appA}).   Interestingly enough,  when the operator  $\Lambda_{p}$ is properly normalized, the resulting functions
converge to $\tilde J_p(x)$.   In the language of dynamical systems theory one could therefore
respectively speak of ``fixed point  of the application defined by the operator''  and ``basin of attraction'' when referring to  $\tilde J_p(x)$ and the set of seed functions which eventually converge to $\tilde J_p(x)$.

As an aside, it is interesting to note that our method is similar in spirit to Neumann's method for tackling integral equations, where the integral operator defining the integral equation is used to generate a function series starting from a zero-th order approximation function \cite{Tricomi}.
In contrast, our method differs significantly from the techniques used in recent works on polynomial approximations for Bessel functions.  For example,  Gross \cite{Gross} exploited the similarity of an integral related to a problem of electromagnetic scattering in a conducting strip grating with an integral representation of $J_p(x)$ to devise a polynomial approximation valid for $J_0(x)$ and $J_1(x)$. Millane and Eads subsequently extended this type of approximation to any  $J_p(x)$ of integer order $p$ ~\cite{MillaneEads}. More recently, Li et al. generalized these results by computing approximations valid for any real $p$ \cite{LiLiGross}. At a later stage, we shall use the results in \cite{LiLiGross} as a reference to test the accuracy of our own approximation.

The present work is organized as follows. Section \ref{sec:predef} is devoted to the introduction of some preliminary definitions. In section \ref{sec:lambdaatt}, the operator $\Lambda_{p}$  is shown to
have an attractor proportional to $\tilde J_p(x)$.  We demonstrate that repeatedly applying
$\Lambda_{p}$ to any non-zero polynomial seed function generates a set of polynomials which
converge to the attractor. In section \ref{sec:funcnorm} we show that the attractor of a slightly modified
operator $\hat\Lambda_{p}$ is $\tilde J_p(x)$ itself. We subsequently study the set  of polynomial
approximations generated by two different seed functions and discuss some numerical results (section \ref{sec:numres}). As shown in section \ref{BamodosFrac}, one of these sets appears naturally in the context
of some fractional diffusion/diffusion-wave problems, thereby describing the long-time decay of the
solutions. Finally, we summarize our main conclusions and briefly outline some avenues for further research in section \ref{sec:concl}.

\section{Preliminary definitions} \label{sec:predef}

For the purpose of finding polynomial approximations to $J_p(x)$, it is convenient to introduce the following set of normalized functions
\be
\tilde J_{p,n}(x)= 2^p p!(z_{p,n} x)^{-p} J_p(z_{p,n}\,x),
\label{tilJJn}
\ee
where $z_{p,n}$ is the $n$-th zero of $J_p(x)$. The function $\tilde J_{p,1}(x)$ will hereafter play a
central role, and hence we shall use the short-hand notation $\tilde J_{p}(x)$ to denote it. Likewise we shall
use $z_{p,1}\equiv z_p$. In terms of the above notation, \eref{tilJJn} can be rewritten as
\be \label{Jpjp} J_p(x)  =\frac{x^p}{2^p
p!} \tilde J_{p}(x/z_p) \ee
for $n=1$. The last equation implies that, if polynomial estimates for $\tilde J_{p}(x)$ are available, they immediately lead to approximations of the same type for $J_p(x)$. From the
series representation of $J_p(x)$, \be J_p(x)=\frac{x^p}{2^p} \sum_{k=0}^\infty \frac{(-1)^k}{k! (k+p)!}
\left(\frac{x}{2} \right)^{2k}, \ee one easily sees that $\tilde J_p(0)=1$.

Let us introduce the following linear integral operator
\be
\label{lambdaDef} \Lambda_{p,n}\left[ f \right]=z_{p,n}^2
\int_x^1\frac{du}{u^{2p+1}} \int_0^udv \,v^{2p+1} f(v)
\ee
acting on the set of functions $\mathcal{C}$ whose elements are those functions for which the double integral performed by $\Lambda_{p,n}$ is well defined.  It is easy to show that any function which is equally singular or less
singular than $x^\alpha$ with $\alpha> \mbox{max}[-2,-2-2p]$ is an element of  $\mathcal{C}$. [We
say that $g(x)$ and $x^\alpha$ are equally singular if $g(x)/ x^\alpha \to \mbox{const.}$ as $x\to 0$;
$g(x)$ is less singular than $x^\alpha$ if $g(x)/ x^\alpha \to 0$ as $x\to 0$].   Note also that any
polynomial function is mapped onto another polynomial function by the operator $\Lambda_{p,n}$.

As shown in section \ref{sec:appA}, the functions $\tilde J_{p,n}(x)$ remain invariant under the
action of the operator  $\Lambda_{p,n}$, i.e., \be \label{invprop} \Lambda_{p,n}\left[ \tilde J_{p,n}(x) \right]=\tilde J_{p,n}(x). \ee In what follows we shall use the short-hand notation $\Lambda_p\equiv \Lambda_{p,1}$ for simplicity. With this notation, one has \be \label{tildejpInva}
\Lambda_{p}\left[ \tilde J_p(x)  \right]= \tilde J_p(x) \ee for the special case $n=1$.

\section{The operator $\Lambda_{p}$ has an attractor proportional to $\tilde J_p(x)$ }
\label{sec:lambdaatt}

We aim to show that \be \label{atteq2} \lim_{n\to\infty}\Lambda^n_{p}[P_s(x)] \propto \tilde J_{p}(x), \ee where $P_s(x)=\sum_{r=0}^s c_r x^r$ is a polynomial of degree $s$. To accomplish this task,
we shall first show that $\Lambda^n_{p}[1]$ has an attractor proportional to $\tilde J_{p}(x)$. Next,
we shall derive an expression for $\Lambda^n_{p}[P_s(x)]$  in terms of $\Lambda^{0}_{p}[1],\ldots, \Lambda^n_{p}[1]$. Finally, we shall argue that the limit $n\to\infty$ of
that expression yields  \eref{atteq2}.

For the sake of simplicity, we shall use the notation $\Lambda_p^{m} [1]=I_m(x)$ in what follows (in this short-hand notation, no explicit reference to the index $p$ is made, which should be inferred from the context). Clearly, the $I_m(x)$'s are polynomial functions of degree $2m$ in $x$. In particular, one of course has $\Lambda^0_{p}[1]=I_0(x)=1$. In order to study the
behaviour of these functions as $m\to\infty$, we now invoke a relation obtained from   (B4), (9) and (11) in reference \cite{divergentSeriesYBA}, namely
\be
\label{ImFou}
I_m(x)=\Lambda_p^m[1]=
\frac{1}{2^{p-1} p!} \sum_{k=1}^\infty \left(\frac{z_p}{z_{p,k}}\right)^{2m}  \;
\frac{z_{p,k}^{p-1}}{J_{p+1}(z_{p,k})}\;  \tilde J_{p,k}(x)
\ee
(for a short demonstration, see \ref{sec:appB}). Note that, as $m$ gets larger, the weight of the
large-$k$ terms in the above sum decreases very rapidly due to the prefactor $z_{p,k}^{-2m}$. In the
limit $m\to\infty$, the only relevant contribution corresponds to the $k=1$ term. Hence, one has
\be
\label{la1tolim}
  I_m(x) \to \zeta_p\;  \tilde
 J_{p}(x), \quad  {\rm as } \quad m\to\infty,
\ee
where the number $\zeta_p$ is
\be
\zeta_p=\frac{1}{2^{p-1} p!}  \frac{z_p^{p-1}}{J_{p+1}(z_p)}.
\ee

Because of the linearity of $\Lambda_p^{m}$, formula \eref{la1tolim} can be rewritten as
\be
\label{defJP1}
 \lim_{m\to\infty} \Lambda_p^{m}[1/\zeta_p ]=  \tilde J_{p}(x).
\ee
Equations \eref{defJP1} and \eref{tildejpInva} provide two remarkable ways of defining  $p$-th order Bessel functions of the first kind by means of the integral operator $\Lambda_p$: (i) as the limit of the feedback process $f_{n+1}(x)=\Lambda_p[f_n(x)]$ with $f_0(x)=1/\zeta_p$, and (ii) as the solution of the fixed point equation $f(x)=\Lambda_p[f(x)]$. These definitions nicely resemble the way in which fractals are defined in terms of the Hutchinson operator \cite{PeitgenJurgensSaupeCF}.  Of course,  equation \eref{la1tolim} also allows one to compute increasingly accurate estimates for the normalized functions $\tilde J_{p}(x)$ and thus for $J_{p}(x)$ by virtue of \eref{Jpjp}.

\subsection{Evaluation of $\Lambda^n_{p}[P_s(x)]$}

A straightforward application of the
operator $\Lambda^n_{p}$ to the monomial $x^r$ gives
 \be \label{lambdaxm}
\Lambda_{p}[x^r]=\frac{1-x^{r+2}}{(2+r)(2+2p+r)}=a_r - a_r x^{r+2} \quad  {\rm for}
\quad p>-1, \; r\ge 0, \ee
where  $a_r=z_p^2/[(2+r)(2+2p+r)]$. It can be  proven by induction
that
\be \label{tlnpxrtilde}
\Lambda^n_{p}[x^r] = \sum_{k=1}^n (-1)^{k-1} b_k(r,p) I_{n-k}(x) +(-1)^n  b_n(r,p) x^{r+2n}. \ee
 In the above expression we have introduced
\be \label{defbkrp}
 b_k(r,p)\equiv \prod_{m=0}^{k-1} a_{r+2m} = z_p^{2k}
 \frac{r!!}{(r+2k)!!}\frac{(2p+r)!!}{(r+2p+2k)!!}
  \ee
as well as the definitions $(2n+1)!!=(2n+1)!/(2^n n!)$  and $(2n)!!=2^n n!$. Note that $b_n(r,p)$ goes rapidly to zero for large $n$: \be \label{coeffdecay} b_n(r,p) \sim (z_p/2)^{2n}/(n!)^2. \ee Applying
$\Lambda^n_{p} $ to an $s$ degree polynomial $P_s(x)=\sum_{r=0}^s c_r x^r$ and using (\ref{defbkrp}) and (\ref{tlnpxrtilde}), one gets
\be \label{tildeLamPs}
\fl  \Lambda^n_{p}[P_s(x)] = \sum_{k=1}^n (-1)^{k-1} I_{n-k}(x) \sum_{r=0}^s c_r b_k(r,p)
+(-1)^n x^{2n} \sum_{r=0}^s c_r b_n(r,p)  x^{r}.
\ee

Next, let us investigate the behaviour of the above expression when $n\to\infty$. First note that,
according to  \eref{coeffdecay}, the coefficients $b_n(r,p)$ roughly go to zero as $(n!)^{-2}$ for large $n$. As a result of this, the second term on the r.h.s. of  \eref{tildeLamPs} becomes negligible with respect to the first one, which is a polynomial of degree $2n-2$ with an independent term (of course, the larger the value of $x$, the larger the value of $n$ necessary to make this term negligible). On the
other hand, it is not difficult to see that the first term of   \eref{tildeLamPs} tends to an expression
proportional to $\tilde J_p(x)$, since (i) according to (\ref{la1tolim}), when $n-k \to \infty$ the
polynomials $I_{n-k}(x)$ tend to the limiting expression $\zeta_p \tilde J_p$, and (ii) those terms in
the sum for which $n-k$ is small, i.e., those terms for which $k$ is large, become negligible due to the fast decay of $b_k(r,p)$ with increasing $k$. We thus conclude that $\Lambda^n_{p}[P_s(x)] \propto \tilde J_p(x)$ when $n\to\infty$.

\section{Generation of $\tilde J_p(x)$ via the normalized operator $\hat \Lambda_{p}$}
\label{sec:funcnorm}

As we have just seen,  when  successively applied to a polynomial, the operator $\Lambda_{p}$
generates  another polynomial which rapidly approaches $\tilde J_{p}(x)$ up to a prefactor. A possible
way to get rid of the prefactor is to divide these polynomials by their value at the origin. The resulting ``normalized'' polynomials becomes equal to one at $x=0$,
which is precisely the value taken by $\tilde J_{p}(x)$ at $x=0$. This prompts us to introduce the ``normalized'' operator
 \be
 \hat \Lambda_{p}[f]=\frac{\Lambda_{p} [f]}{\Lambda_{p} [f]_{x=0}}
\ee acting on the subset of functions $f \in {\cal C}$ for which $\Lambda_{p} [f]_{x=0}$ (that is,
$\Lambda_{p} [f]$ evaluated at $x=0$) is non-zero.
In particular, using  \eref{tildeLamPs}, we have 
\begin{equation}
\label{hatLamPsTilde} 
\fl \hat\Lambda^n_{p}[P_s(x)] = \frac{\sum_{k=1}^n (-1)^{k-1} I_{n-k}(x)  \sum_{r=0}^s c_r b_k(r,p) +(-1)^n x^{2n} \sum_{r=0}^s c_r b_n(r,p)  x^{r} } {\sum_{k=1}^n (-1)^{k-1} I_{n-k}(0)  \sum_{r=0}^s c_r b_k(r,p)} \end{equation}
 and $\hat \Lambda^n_{p}[P_s(x)]\to \tilde J_p(x)$ as $n\to\infty$.  In view of the above derivation, one expects that any well-behaved $f(x)$  (in the sense that it can be approximated arbitrarily well by a polynomial) also converges to the same attractor, i.e., \be \label{fundeq} \hat \Lambda^n_{p}[f(x)] \to \tilde J_p(x), \quad n\to\infty. \ee

The iteration of the operator  $\hat \Lambda_{p}$  will allow us to generate families of approximations to
$J_p(x)$ by using different seed functions $f_0(x)$. Each of the seed functions $f_0(x)$ leads to a series of
functions $\{f_0, f_1,f_2, \ldots\}$, with $f_n=\hat \Lambda^n_{p}[f_0]$, that converges to $\tilde J_p(x)$ as $n\to\infty$.
One expects that the convergence to the attractor becomes faster as the initial condition gets closer
to $\tilde J_p(x)$. On the other hand, from a practical point of view it would be desirable that the chosen
seed function $f_0$ is simple enough to ensure that the integrals resulting from the iteration of $\hat
\Lambda_p$ are known and easy to calculate analytically. Obvious candidates are polynomials.  Surely, the most simple initial function is  $f_0(x)=1$. This function gives rise to the set
$\{f_n(x)\equiv {\rm Ba}^{(p)}_n(x)\} $, defined as
\be \label{Badef}
  {\rm Ba}^{(p)}_n(x) \equiv
\hat  \Lambda^{(n)}_{p}[1] =\frac{I_n(x)} {I_n(0)}. \ee
The initial condition $f_0(x)=P_s(x)=1$ is a particular case of a polynomial function for which
 $c_0=1$ and $c_r=0 \,(r>0)$. Inserting this expression into \eref{tildeLamPs} and recalling the
 definition of $I_n(x)$ one gets
\be \label{recrelIn}I_n(x) =  \sum_{k=1}^n (-1)^{k-1}b_k(0,p) I_{n-k}(x)  +(-1)^n b_n(0,p) x^{2n}, \ee where  $  b_k(0,p) =  z_p^{2k} p!/[ 2^{2k} k!
 (p+k)!]$ [cf.~\eref{defbkrp}]. Inserting this expression into \eref{recrelIn} one finds
\be  \label{Inx}
  I_n(x) =  \sum_{k=1}^n (-1)^{k-1} \frac{z_p^{2k} p!}{ 2^{2k} k! (p+k)!} I_{n-k}(x)
  +(-1)^n \frac{z_p^{2n} p!}{ 2^{2n} n! (p+n)!} x^{2n}.
\ee The first few polynomials $ {\rm Ba}^{(p)}_n(x)$ computed from  (\ref{Badef}) and (\ref{Inx})
read
 \begin{eqnarray*} {\rm Ba}^{(p)}_0(x) =&1,\\  {\rm Ba}^{(p)}_1(x) =&1-x^2,\\  {\rm Ba}^{(p)}_2(x)
=&1-\frac {2 (p + 2)} {p + 3} x^2 +\frac {p + 1} {p + 3} x^4,\\ {\rm Ba}^{(p)}_3(x) =&1-\frac {3 (p + 3)^2}
{p^2 + 8 p + 19} x^2 +\frac {3 (p + 1) (p + 3)} {p^2 + 8 p + 19} x^4 -\frac {(p + 1)^2} {p^2 + 8 p + 19}
x^6.
\label{fam1primeros}
\end{eqnarray*}

As a test of the present method, we shall also investigate the behaviour of another set of functions
$  {\rm Be}^{(p)}_n(x)$
generated from the  $f_0(x)=1-x$. In this case, from \eref{tildeLamPs} and the definition of $\hat \Lambda^n_{p}$ one finds
 \be 
 \label{hatLamP1}
\fl \hat \Lambda^n_{p}[1-x] = \frac{\sum_{k=1}^n (-1)^{k-1} I_{n-k}(x)  [ b_k(0,p)-b_k(1,p)] +(-1)^n x^{2n}  [ b_k(0,p)- b_k(1,p) x]   } {\sum_{k=1}^n (-1)^{k-1} I_{n-k}(0)
[b_k(0,p)-b_k(1,p)] }. 
\ee 
The corresponding polynomials can be easily evaluated via the equations
 \eref{Inx} and \eref{defbkrp}. The first few polynomials are given below:
\begin{eqnarray*}  {\rm Be}^{(p)}_0(x)=&1-x,\\  {\rm Be}^{(p)}_1(x)=& 1-\frac {6 p + 9} {2 p + 5} x^2 +\frac
{4 (p + 1)} {2 p + 5} x^3,\\ {\rm Be}^{(p)}_2(x)=&1-\frac {10 (p + 2) (2 p + 5)^2} {3 \left (4 p^3 + 36 p^2 +
115 p + 113 \right)}  x^2 +\frac {5 (p + 1) \left (4 p^2 + 16 p + 15 \right)} {4 p^3 + 36 p^2 +
   115 p + 113}  x^4\\
   & - \frac {32 (p + 1)^2 (p + 2)} {3 \left (4 p^3 + 36 p^2 + 115 p +
     113 \right)} x^5 .
          \label{fam1primeros}
\end{eqnarray*}

\section{Numerical results} \label{sec:numres}

\begin{figure}
 \begin{center}
         \includegraphics[width=0.75\columnwidth]{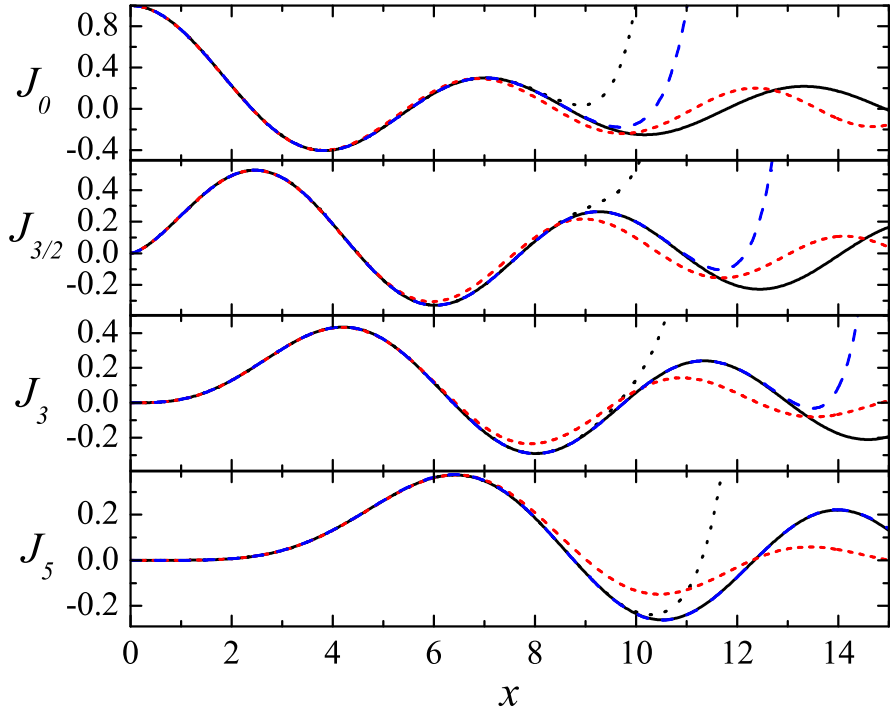}
\end{center}
 \caption{Comparison of  $J_p(x)$ (solid line) with the LLG polynomial approximation of order
$2\times 10+p$ \cite{LiLiGross}, $L_{10}^{(p)}$, (short dashed line), the truncated Taylor series (dotted line) and the polynomial approximation ${\rm Ba}_p^{(10)}(x)$ of $ \tilde J_{p}(x)$ (dashed line) [see (\ref{Badef}) and (\ref{Inx})] for $p=0, 3/2, 3, 5$.
In the lower panel the line corresponding to our approximation for $J_5(x)$ lies on top of the exact one. } \label{fig:ComparcionconLiLiGross} \end{figure}

Li, Li and Gross (LLG) proposed in Ref. \cite{LiLiGross} an approximation to $J_p(x)$ based on its integral representation.  They obtained an infinite series whose truncation leads to the following polynomial approximation of $J_p(x)$,
\be
L_n^{(p)}= \sum_{m=0}^n (-1)^m \frac{ n^{1-2 m}  (m+n-1)! }{m! (n-m)! \Gamma (m+p+1)} \, \left(\frac{x}{2}\right)^{2 m+p}
\ee
thereby providing an alternative approximation to truncated Taylor series. Figure 1 in \cite{LiLiGross} depicts a comparison of $J_p(x)$ with the LLG polynomial approximation of degree $2\times 10+p$ and with the Taylor series truncated to the same degree for  $p=0, 3/2, 3, 5$.
 In figure \ref{fig:ComparcionconLiLiGross} of the present paper we have
superposed their results with our own results based on the approximation obtained by making the replacement $\tilde J_p(x/z_p)\to{\rm Ba}_{10}^{(p)} (x/z_p)$ in Eq. \eref{Jpjp}.
We see that our approximation performs very well over a larger range of  $x$ values than the LLG polynomial approximation and the truncated Taylor series of the same order.  Even though the LLG polynomial approximation oscillates around $J_p(x)$ over a larger $x$ interval than the other two approximations, it starts deviating from the exact curve at smaller values of $x$. In all cases, deviations from $J_p(x)$ are shifted to larger $x$ values with increasing order $p$. In this sense, all three approximations become better with increasing $p$.
Regarding the computational efficiency, the CPU time employed for calculating the polynomial approximations obviously depends on their order $n$. For example, when using the program \textit{Mathematica}, the CPU time necessary for computing the curves in the top panel of figure \ref{fig:ComparcionconLiLiGross} (corresponding to $n=10$) is roughly the same for each of the three polynomial approximations depicted; this time is roughly one half of the time required to compute the curves corresponding to $n=20$ and, also, one half of the time required when using directly the Bessel function algorithm.

\begin{figure} \begin{center}
          \includegraphics[width=0.75\columnwidth]{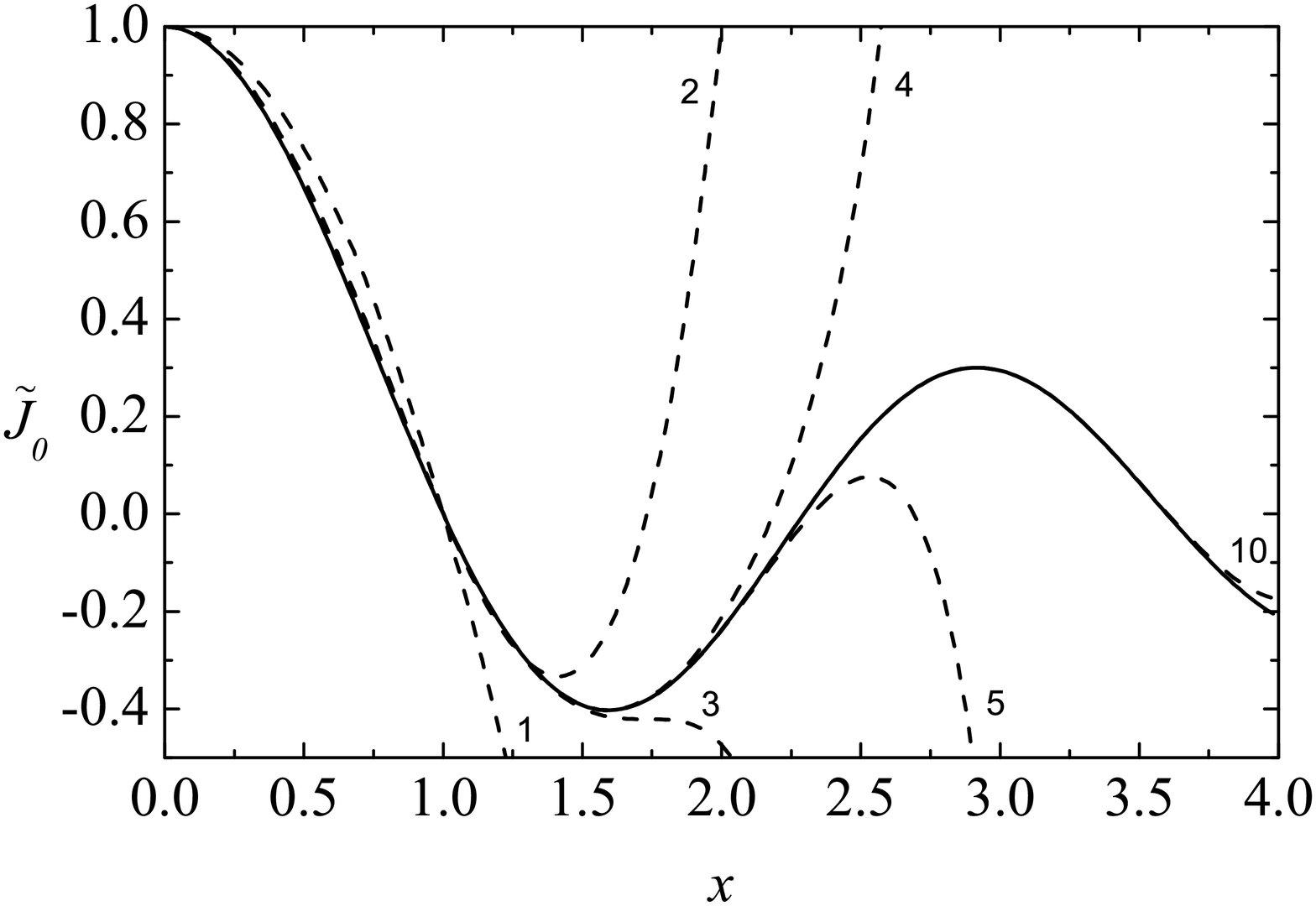}
\end{center} \caption{Comparison between $\tilde J_0(x)$  (solid line) and ${\rm Ba}^{(0)}_n(x)$ with
$n=1, 2, 3, 4, 5, 10$. The value of $n$ corresponding to each line is shown. } \label{fig:Ba0Comp}
\end{figure}

\begin{figure} \begin{center}
          \includegraphics[width=0.75\columnwidth]{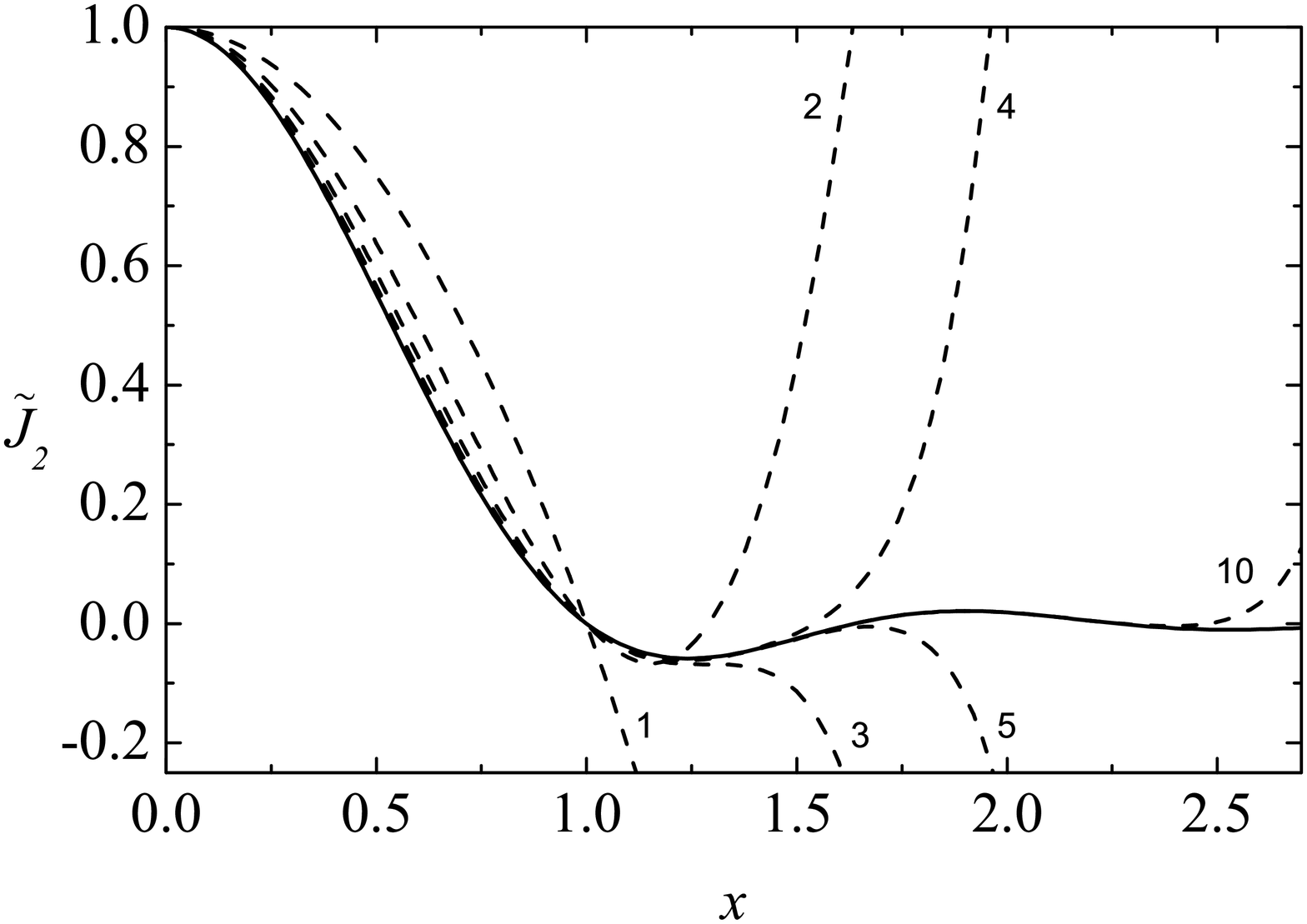}
\end{center} \caption{Comparison between $\tilde J_2(x)$  (solid line) and ${\rm Ba}^{(2)}_n(x)$ with
$n=1, 2, 3, 4, 5, 10$.  The value of $n$ corresponding to each line is shown.} \label{fig:Ba2Comp}
\end{figure}

In figure \ref{fig:Ba0Comp} [figure \ref{fig:Ba2Comp}] we compare   $\tilde J_0(x)$  [$\tilde  J_2(x)$] with
the polynomials ${\rm Ba}_n^{(0)}(x)$ [${\rm Ba}_n^{(2)}(x)$]  for $n=1, 2, 3, 4, 5, 10$. For fixed values
of $n$ and $x$ the approximation is more accurate for $p=0$ than for $p=2$. For other $p$ values we
have checked that, in general, the accuracy of our approximation for $\tilde J_p(x)$ decreases with
increasing $p$, as opposed to the behaviour observed for $J_p(x)$ (see figure
\ref{fig:ComparcionconLiLiGross}). Clearly, the different behaviour is related to the rescaling
introduced by   \eref{Jpjp}.

\begin{figure} \begin{center} \includegraphics[width=0.75\columnwidth]{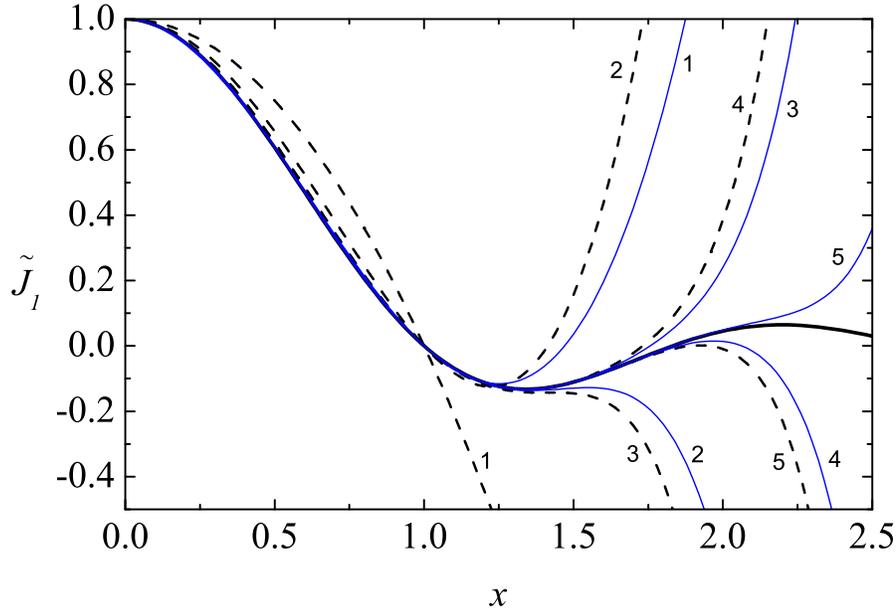} \end{center}
\caption{Comparison between $\tilde J_1(x)$  (thick solid line) and ${\rm Ba}^{(1)}_n(x)$ (dashed lines)
and ${\rm Be}^{(0)}_n(x)$ (thin solid lines)  for  $n=1, 2, 3, 4,5$.  The value of $n$ corresponding to each line is shown.  } \label{fig:Ba1Be1Comp} \end{figure}

Finally, we proceed to compare $\tilde J_1 (x)$ with the polynomial approximations
${\rm Ba}^{(1)}_n(x)$ and ${\rm Be}^{(1)}_n(x)$  for $n=1, 2, 3, 4, 5, 10$. This comparison is shown in
figure \ref{fig:Ba1Be1Comp}. Not surprisingly, the family ${\rm Be}^{(1)}_n(x)$ performs better than ${\rm Ba}^{(1)}_n(x)$ since  it starts from the seed function $f_0(x)=1-x $ which is a better approximation to $\tilde J_1 (x)$ than $f_0(x)=1$.

\section{The polynomials ${\rm Ba}^{(p)}_n(x)$ describe fractional diffusive and oscillatory modes}
\label{BamodosFrac}

As shown in  \cite{divergentSeriesYBA}, the polynomials ${\rm Ba}^{(p)}_n(x)$ can be used to
express the long-time decay modes in some fractional diffusion problems in a $d$-dimensional sphere. In
what follows we shall demonstrate that the ${\rm Ba}^{(p)}_n(x)$'s also appear in problems whose
solution is given by the fractional diffusion-wave equation  subject to the same geometry and boundary
conditions (note, however, that for diffusion-wave problems one must additionally specify the initial
velocity).  Take, for instance, the problem described by the equation \cite{Mainardi96}
\be
\label{frac-diff-wav} \frac{d^\gamma c(r,t)}{dt^\gamma}=K_\gamma \nabla^2 c(r,t),
\ee
where
$K_\gamma$ is a coefficient and the operator $d^\gamma/dt^\gamma$
 denotes the Caputo fractional derivative of order $\gamma$ \cite{Mainardi96,MainardiPagnini03,Podlubny}:
\be  \frac{d^\gamma f(t)}{dt^\gamma}= \frac{1}{\Gamma(n-\gamma)} \int_0^t
\frac{f^{n}(\tau)}{(t-\tau)^{\gamma+1-n} } \, d\tau, \quad n-1<\gamma<n
\ee
with $n$ an integer. In
\cite{divergentSeriesYBA}, problem B in Section III, the long-time solution of \eref{frac-diff-wav} corresponding to the
boundary condition $c(R,t)=0$ and the initial condition $c(r,0)=c_0$ was discussed for the case
$0<\gamma\le 1$.  Physically, $c(r,t)$ represents the decay of a homogeneous
initial particle concentration inside a hyperspherical volume with an absorbing boundary of radius $R$.
The solution consists of a series of decay modes whose spatial part is expressible in terms of the
polynomials ${\rm Ba}^{(p)}_n(x)$ via a recursion relation essentially identical with \eref{Inx}.

In the range of values $1<\gamma < 2$,    \eref{frac-diff-wav} becomes a diffusion-wave equation \cite{Mainardi96, Mainardi96b, GorenfloLuch, Hanygad}.
For $c(r,0)=c_0$ and $dc(r,t)/dt=0$ at $t=0$ the solution reads
 \begin{equation}
  \label{solejemplo2Sub}
  c(r,t)/c_0=  2(r/R)^{-\eta}\sum_{j=1}^{\infty}\frac{J_\eta(z_{\eta,j}\, r/R)}
{z_{\eta,j} J_{\eta+1}(z_{\eta,j})} E_{\gamma} \left[-(z_{\eta,j}/R)^2 K_{\gamma}t^{\gamma}\right]
\end{equation} with $\eta=d/2-1$, which is also the solution for $0<\gamma\le 1$ (fractional diffusion equation) when $c(r,0)=0$ . The solution is obtained by separation of variables and subsequent use of the
relation \cite{Mainardi96,MainardiPagnini03}
 \be
 \frac{d^\gamma}{dt^\gamma}E_{\gamma} \left[-\omega^\gamma t^{\gamma}\right]=
 -\omega^\gamma E_{\gamma} \left[-\omega_\gamma t^{\gamma}\right]
 \ee
 for the derivative of the Mittag-Leffler function $E_\gamma[\cdot]$.
Inserting the asymptotic expansion of the Mittag Leffler functions \be
E_\gamma\left[-z\right]=\sum_{m=1}^\infty\frac{(-1)^{m+1}}{\Gamma(1-m\gamma)} z^{-m}, \qquad
z\to\infty \ee for $0<\gamma<1$ and $1<\gamma< 2$ into   \eref{solejemplo2Sub} and grouping the
terms with the same power of $t$, one finds
 \be
 \label{solAsinejemplo2SubA}
  c(r,t)/c_0=\sum_{m=1}^{\infty}\frac{(-1)^{m+1} I_m(0) }{z_p^{2m} \Gamma(1-m\gamma)}
\left(\frac{R^2}{K_\gamma t^\gamma}\right)^m    {\rm Ba}_m^{(\eta)}(r/R) \ee for $0<\gamma<1$ and
$1<\gamma< 2$ (in the above equation $I_m(0)$ must be evaluated for $p=\eta$). Thus, the spatial
dependence of the long-time solution can be expressed in terms of a suitable superposition of the
``fractional modes'' ${\rm Ba}_m^{(\eta)}(x)$. The above fractional solution extends the classical solution (cases $\gamma=1,2$), where the spatial modes are proportional to Bessel functions.  In other words, the role of
the $m$-th $d$-dimensional normal mode $ J_{\eta}(z_{\eta,m} r/R)/(r/R)^{\eta}$ in the normal diffusion
equation ($\gamma=1$) and wave equation ($\gamma=2$) is played by the $m$-th $d$-dimensional
fractional mode ${\rm Ba}_m^{(\eta)}(r/R)$ in the fractional diffusion equation ($0<\gamma<1$) and
fractional diffusion-wave equation ($1<\gamma<2$) : \be \frac{J_{\eta}(z_{\eta,m} r/R)}{(r/R)^{\eta} }
\longleftrightarrow {\rm Ba}_m^{(\eta)}(r/R). \ee

\section{Summary and Outlook} \label{sec:concl}

 The present work deals with a novel method to obtain polynomial approximations to Bessel functions of the first kind. In some cases the obtained polynomials turn out to be more accurate than truncated Taylor series and the  polynomials in  \cite{LiLiGross} over a wide range of parameter values. We have seen that the polynomials  ${\rm Ba}_n^{(p)}(x)$ generated by the initial function $f_0(x)=1$ are  interesting in their own right, as they represent the spatial modes describing the long-time behaviour of certain solutions of fractional diffusion and diffusion-wave equations.

From a computational point of view, a nice property of our method is that for fixed $x$ and $p$ the distance $ \left|{\rm Ba}_{n+1}^{(p)}(x)-{\rm Ba}_n^{(p)}(x)\right|$ between successive
approximations decreases rapidly with increasing iteration number $n$, and one can set a threshold value  below which, for practical purposes, convergence to the corresponding Bessel function may be considered to have taken place.

Integral operator methods developed along similar lines might be able to provide alternative polynomial approximations in the range $p\le-1$. Of course  our solution is also valid for negative integer values of $p$ by virtue of the relation $J_{-p}(x)= (-1)^p J_p(x)$ for $p$ integer.
Such techniques might also be useful for other kinds of Bessel functions (e.g. Bessel functions of the second and the third kind). An open question is whether similar iterative methods can be applied to generate polynomial approximations to other special functions associated with ordinary differential equations.

\ack Financial support from the Ministerio de Educaci\'on  y Ciencia through Grant No. FIS2007-60977 (partially financed with FEDER funds) and by the Junta de Extremadura (Spain) through Grant Nos. GRU09038 and GRU10158 is gratefully acknowledged.

\appendix

\section{Invariance of $\tilde J_{p,n}(x)$ under the action of $\Lambda_{p,n}$}
\label{sec:appA}

The starting point is the differential equation fulfilled by the Bessel functions \be y^2 \frac{d^2
J_p(y)}{dy^2}+y\frac{d J_p(y)}{dy}+\left(y^2-p^2 \right)J_p(y)=0. \ee Using the transformation
$y=z_{n,p}\,v$ in the above equation gives \be v^2 \frac{d^2 J_p(z_{p,n}\,v)}{dv^2}+v\frac{d
J_p(z_{p,n}\,v)}{dv}+\left(z_{p,n}^2\,v^2-p^2 \right)J_p(z_{p,n}\,v)=0. \ee Multiplying by
$v^{p-1}/z_{p,n}^2$ and rearranging terms we get
\be
\fl -v^{p+1}J_p(z_{p,n}\,v)=\frac{v^{p+1}}{z_{p,n}^2}\frac{d^2
J_p(z_{p,n}\,v)}{dv^2}+\frac{v^p}{z_{p,n}^2}\frac{d
J_p(z_{p,n}\,v)}{dv}-\frac{p^2\,v^{p-1}}{z_{p,n}^2}J_p(z_{p,n}\,v). \ee The above equation can be
rewritten as follows
\be -v^{p+1}J_p(z_{p,n}\,v)=\frac{1}{z_{p,n}^2}\frac{d}{dv}
v^{2p+1}\frac{d}{dv}\left[v^{-p}J_p(z_{p,n}\,v)\right]. \ee
Integrating between $0$ and $u$ we get \be
\label{inteq} -\frac{z_{p,n}^2}{u^{2p+1}}\,\int_0^u
dv\,v^{p+1}J_p(z_{p,n}\,v)=\frac{d}{du}\left[u^{-p}J_p(z_{p,n}\,u)\right], \ee where we have used \begin{eqnarray}
\lim_{v\to 0}v^{2p+1}\frac{d}{dv}\left(v^{-p}J_p(z_{p,n}\,v)\right)= \lim_{v\to
0}\left[v^{p+1}\frac{dJ_p(z_n\,v)}{dv}-pv^p J_p(z_{p,n}\,v)\right] \nonumber \\
= -z_{p,n} \lim_{v\to 0}
v^{p+1}J_{p+1}(z_{p,n}\,v)=0 \quad (p>-1). 
\end{eqnarray} Integrating once again \eref{inteq} between $1$ and $x$
and using $J_p(z_{p,n})=0$ we get \be z_{p,n}^2\int_x^1 \frac{du}{u^{2p+1}}\int_0^u
dv\,v^{2p+1}\,v^{-p}J_p(z_{p,n}\,v)=x^{-p}J_p(z_{p,n}\,x). \ee Finally, multiplying the above equation by
$2^p\,p!z_{p,n}^{-p}$ yields   \eref{invprop}.

\section{Proof of formula \eref{ImFou} }
\label{sec:appB}

The Fourier-Bessel expansion of a function $g(x)$ is given by
$ g(x)= \sum_{k=1}^{\infty}c_k J_{p}(z_{p,k} x)$
with
\begin{equation}
\label{ckdef}
    c_k=\frac{2}{J_{p+1}^2(z_{p,k})}\int_0^1 x g(x)  J_{p}(z_{p,k} x) dx.
\end{equation}
Therefore, \eref{ImFou} is equivalent to the statement that the coefficients of the Fourier-Bessel expansion of  $x^p I_n(x)$ are
\be
\label{ckpn}
c_k \equiv c(k,p,n)=
2   \left(\frac{z_p}{z_{p,k}}\right)^{2n}  \;
\frac{z_{p,k}^{p-1}}{J_{p+1}(z_{p,k})}\;.
\ee
We are going to prove this equation by induction. To start with, it is well-known \cite{Watson}  that the coefficients of the Fourier-Bessel expansion of  $x^p I_0(x)=x^p$ are
\be
c(k,p,0) =\frac{2 z_{p,k}^{-1}}{ J_{p+1}(z_{p,k})}.
\ee
This justifies \eref{ckpn} for $n=0$. Therefore, to prove equation \eref{ImFou} is equivalent to prove that $c(k,p,n+1)=(z_p/z_{p,k})^2 c(k,p,n)$. By using \eref{ckdef} with $g(x)=x^p I_{n+1}(x)$ and  making the substitution $y=z_{p,k} x$ inside the integral one sees that
\be
\label{ck4}
c(k,p,n+1)=\frac{2}{z_{p,k}^{p+2} J_{p+1}^2(z_{p,k})}\int_0^{z_{p,k}}  y^{p+1} J_{p}(y) I_{n+1}(y/z_{p,k}) dy.
\ee
Using $y^{p+1} J_{p}(y) =d[y^{p+1} J_{p+1}(y)]/dy$, and integrating by parts one gets
\begin{equation}
\label{ck5}
c(k,p,n+1)=-\frac{2}{z_{p,k}^{p+2} J_{p+1}^2(z_{p,k})}\int_0^{z_{p,k}}  y^{p+1} J_{p+1}(y) \frac{d}{dy}I_{n+1}(y/z_{p,k}) dy
\end{equation}
as the boundary terms vanish. But, from \eref{lambdaDef},
\be
I_{n+1}(x)=z_{p}^2
\int_x^1\frac{du}{u^{2p+1}} \int_0^udv v^{2p+1}I_{n}(v) .
\ee
Inserting this expression into \eref{ck5}, using the relation $J_{p+1}(y)/y^p=-d[J_p(y)/y^p]/dy$  and integrating by parts one gets
\be
c(k,p,n+1)=\frac{2 z_k^2}{z_{p,k}^{p+4} J_{p+1}^2(z_{p,k})}\int_0^{z_{p,k}}  y^{p+1} J_{p}(y) I_{n}(y/z_{p,k}) dy
\ee
as the boundary terms vanish. Comparing this result with the expression of $c(k,p,n)$  given by \eref{ck4}  one sees that $c(k,p,n+1)=(z_k^2/z_{p,k}^{2})c(k,p,n)$, which is just the result we aimed to prove.

\end{document}